\documentclass[aip, amsmath, amssymb, reprint]{revtex4-2}

\usepackage[utf8]{inputenc}
\usepackage[T1]{fontenc}
\usepackage{graphicx}
\usepackage{siunitx}

\draft 

\begin{document}

\title{A Highly Drift-stable Atomic Magnetometer for Fundamental Physics Experiments}
\author{M. Rosner}
\email{martin.rosner@tum.de}
\affiliation{Physikdepartment, Technische Universität München, 85748 Garching, Germany}
\author{D. Beck}%
\affiliation{Department of Physics, University of Illinois, Urbana-Champaign, Illinois 61801, USA}%
\author{P. Fierlinger}
\author{H. Filter}
\author{C. Klau}
\author{F. Kuchler}
\author{P. R\"o{\ss}ner}
\author{M. Sturm}
\author{D. Wurm}
\affiliation{Physikdepartment, Technische Universität München, 85748 Garching, Germany}
\author{Z. Sun}
\affiliation{Laboratory for Space Environment and Physical Sciences, Harbin Institute of Technology, Harbin, China}

\date{\today}

\begin{abstract}
	We report the design and performance of a non-magnetic drift stable optically pumped cesium magnetometer
	with a measured sensitivity of 35~fT at 200~s integration time and stability below 50~fT between 70~s and 600~s. To our knowledge this is the most stable magnetic field measurement to date.
	The sensor is based on the nonlinear magneto-optical rotation effect: in a Bell-Bloom configuration a higher order polarization moment 
 (alignment) of Cs atoms is created with a pump laser beam in an anti-relaxation coated Pyrex cell under vacuum, filled with Cs vapor at room temperature.
	The polarization plane of light passing through the cell is modulated due the precession of the atoms in an external magnetic field of  \SI{2.1}{\micro\tesla}, used to optically determine the Larmor precession frequency.
	Operation is based on a sequence of optical pumping and observation of freely precessing spins at a repetition rate of 8~Hz. 
	This free precession decay readout scheme separates optical pumping and probing and thus ensures a systematically highly clean measurement.
	Due to the residual offset of the sensor of $<$~15~pT together with the cross-talk free operation of adjacent sensors, this device is uniquely suitable for a variety of experiments in low-energy particle physics with extreme precision, here as highly stable and systematically clean reference probe in search for time-reversal symmetry violating electric dipole moments.
	
\end{abstract}


\maketitle 
Precision measurements in particle physics at very low energies recently gained increased attention, as high energy accelerators do not see evidence of new physics to explain big open questions like the nature of dark matter, dark energy or baryon asymmetry in the Universe (BAU).
A  prominent example is the search for the time-reversal symmetry violating electric dipole moment of the neutron (nEDM) \cite{chupp_2019}: if observed, it is a manifestation of new physics needed to explain the BAU.
Our work is motivated specifically by the PanEDM experiment \cite{wurm_2019}, which relies on Ramsey-type magnetic resonance measurements in ultra-low magnetic fields \cite{ramsey_1950} under extremely well controlled conditions\cite{altarev_2015}.
The requirement for the stability of the shape of the magnetic field is determined by the so-called geometric phase effect\cite{pendlebury_2004}.
To reconstruct the magnetic field geometry the neutron storage chambers will be surrounded by an array of optically pumped magnetometers.
To fulfill the requirements for the PanEDM experiment at least eight sensors are needed, spaced at a distance of approximately 60~cm and achieving a resolution of 50~fT over one measurement cycle of 250~s.

Recently, atomic magnetometers \cite{budker_2007} have started replacing SQUID magnetometers in many applications.
These sensors are typically optimized for applications in biomagnetism and medical imaging at frequencies between 1-100~Hz, at noise levels of about $\SI{15}{\femto\tesla\per\sqrt{\hertz}}$ and without the need of applied holding fields\cite{shaw_2013}.
However, most sensors are intrinsically magnetic, generate RF fields or are not stable against various drifts, e.g. due to currents inside the actual sensor head, temperature variations or laser power dependent light-atom interactions. Additionally, their capability of measuring absolute magnetic fields is limited.\\
Here we present a reliable sensor featuring high precision and accuracy at long integration times. Several sensors can be operated simultaneously in an array without cross-talk.
In the following, the principle of operation, the design of the sensor, its operation and systematic studies in the context of long integration times are discussed.
The basic steps of a magnetic field measurement are:

Optical pumping: Cs atoms in an anti-relaxation coated vapor cell under vacuum at room temperature are polarized by using a linearly polarized pump laser beam tuned to the D2 transition.
The generated population difference of the Zeeman sublevels yields an alignment state of two-fold symmetry.
This method, called alignment pumping, suppresses vector light-shifts and results in the absence of a net magnetization of the polarized vapor, which in turn avoids crosstalk among the sensors in close proximity.

\begin{figure}[ht]
	\centering
	\includegraphics[width=8.6cm]{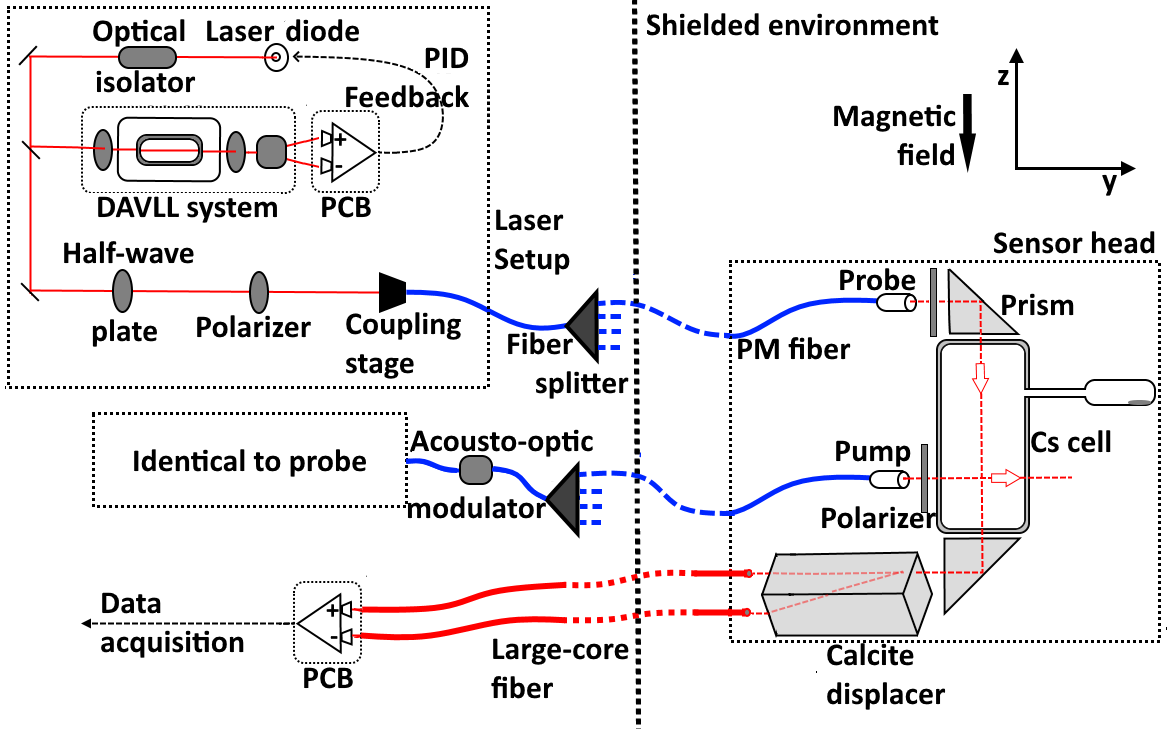}
	\caption{Magnetometer scheme (showing a single sensor head for simplicity): The fiberized, non-magnetic sensor head is placed in the magnetic field, where Cs atoms in the vapor cell are optically pumped by a pulsed laser beam. The rotation of the polarization axis of a much weaker, continuous probe beam is detected as the difference of separated polarization components.}
	\label{fig:sensor_assembly}
\end{figure}

Spin precession: The aligned spins of the Cs atoms precess coherently about the axis of the external magnetic field $\vec{B}$. The corresponding precession frequency, i.e. the Larmor frequency $\omega_\text{L}$, depends on the magnitude of $\vec{B}$ due to the energy shift of the Zeeman sublevels. Due to the two-fold symmetry of the alignment state the observed rotation frequency is $2\omega_\text{L}$. 
Detection: With the polarized vapor showing an optically active behavior, the magnetic field information at any point in time can be extracted optically, in our case through a weak, linearly polarized laser beam. 
The entire procedure is based on nonlinear magneto-optic rotation (NMOR) \cite{budker_2002}.
As the probe laser passes through the atomic vapor, its polarization axis is rotated and finally acquires a sinusoidal modulation at $2\omega_L$. 
With the gyromagnetic ratio\cite{arimondo_1977} $\gamma_\text{Cs} \approx \SI{3.5}{\kilo\hertz\per\micro\tesla}$ the frequency-to-field conversion factor is $\approx \SI{7}{\kilo\hertz\per\micro\tesla}$. 

Next to the sensitivity target of $\SI{50}{\femto\tesla}$ over 250~s integration time additional design criteria of the sensor are:  (i) scalability to enable operation in a sensor array; (ii) a diameter of $\leq$~55~mm of the sensor head; (iii) fully non-metallic and non-magnetic components and (iv) remote operation of the sensor head.

The Cs vapor is contained in a Pyrex glass cell of 10~mm inner diameter and 30~mm length. 
The vapor density is typically $\sim 10^{11}\;\text{cm}^{-1}$ at the sensor's operating temperature of 300~K.
The cells are prepared with a special paraffin coating \cite{graf_2005}, preserving the atomic polarization state for more than $10^4$ atom-wall collisions.
We observed transverse spin lifetimes $T_2$ of 50-70~ms, with typical probe beam powers of \SIrange{3}{8}{\micro\watt} and longitudinal spin lifetimes $T_1$ of $>$150~ms.
The cell is placed with its axis parallel to the magnetic field, $\vec{B}\,||\,\hat{z}$ and mounted in a 3D-printed holder made from  photopolymer plastic using a Formlabs Form 2 3D printer.
The pump beam propagates across the diameter of the cell, $\vec{k}_\text{pump}\,||\, \hat{y}$ with linear polarization oriented perpendicular to the magnetic field, $\vec{\epsilon}_\text{pump}\,||\,\hat{x}$. 
The probe beam, propagating along the axis of the cell, $\vec{k}_\text{probe}\,||\,\hat{z}$, can be linearly polarized in the $\hat{x}$ or $\hat{y}$ direction. 
A simplified schematic of laser optics and sensor head is shown in Fig. \ref{fig:sensor_assembly}.
Two DFB \footnote{Toptica Eagleyard, EYP-DFB-0852-00150-1500-TOC03-0005} laser diodes provide the light for pump and probe beams. Stabilization of the laser frequency is realized using a dichroic atomic-vapor laser lock (DAVLL) \cite{cheron_1994a}.
Modulation of the pump beam intensity is achieved by an acousto-optical modulator (AOM).
The light is coupled into 8~m long polarizing fibers\footnote{Fibercore, Zing HB830Z(5/80)}, which are mounted in the sensor head and terminated by gradient-index lenses. 
\begin{figure}[t]
	\centering
	\includegraphics[width=8cm]{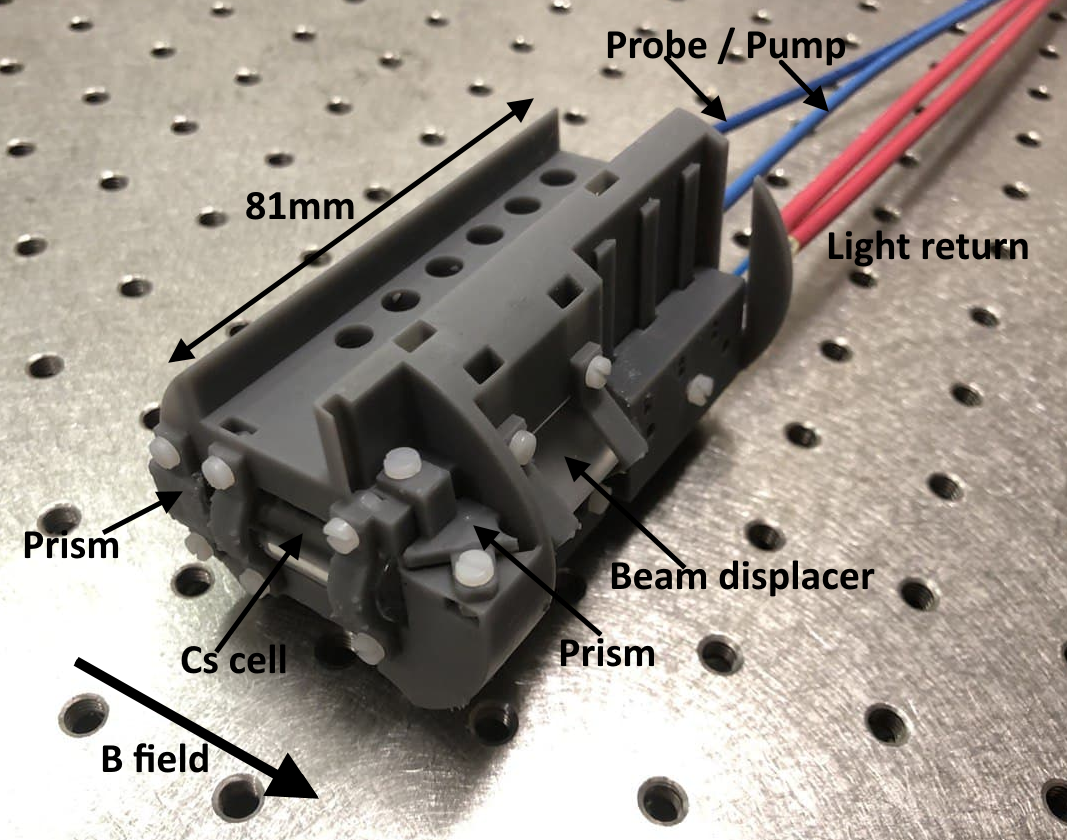}
	\caption{Photograph of the 3D-printed sensor head with main components indicated.}
	\label{fig:sensorhead}
\end{figure}
The pump beam fiber is placed normal to the axis of the Cs cell with a cleanup linear polarizer between fiber and glass cell. %
Precision of mechanical alignment is defined by the precision of the 3D printed sensor mount, which is of order \SI{100}{\micro\metre}.
The probe beam light passes through a cleanup linear polarizer, oriented in the $\hat{x}$ direction, and is reflected by a prism into the cell (c.f. Fig. \ref{fig:sensor_assembly}).
All optical interfaces up to the cell use index matching gel \footnote{Thorlabs G608N3 Index Matching Gel} to minimize reflections, which significantly enhances transmission.
After interacting with the Cs atoms, the probe beam is reflected into a calcite beam displacer by an anti-relaxation coated prism.
The optical axis of the beam displacer is oriented at $45^\circ$ to the $\hat{x}$ axis, separating the beam into two equal intensity components. 
The ordinary ray passes straight through the calcite crystal, the extraordinary ray is displaced by 2.7~mm, producing two parallel beams that enter a pair of 10 m long multi mode large core fibers joined to the crystal with optical index matching gel. The return light is propagated back onto two photodiodes mounted on a PCB board\cite{patterson_2015}, which generates the read out signal as the difference of the two probe beam polarization components. The transfer function of the circuit is optimized to reach a 3 dB cut-off at $\approx \SI{25}{\kilo\hertz}$.
Due to the screw-mounted optics (see Fig. \ref{fig:sensorhead}) and use of non-permanent optical gel, all components can be replaced with low effort if necessary. The Cs cell of a sensor head was successfully replaced on site without loss of performance.\\
Several measures are taken to improve long-term stability. Polarizing optical fibers are used for pump and probe light. In addition, cleanup polarizers inside the sensor suppress any remanent non-linear components and polarization noise in both beams. Splitting the signal into polarization components by positioning the calcite displacer directly after the cell avoids possible polarization noise in outgoing fibers. The DAVLL system is temperature stabilized and pumping frequencies as well as sampling rates are synchronized to a common precision rubidium frequency standard.
 \begin{figure}[t]
	\centering
	\includegraphics[width=9cm]{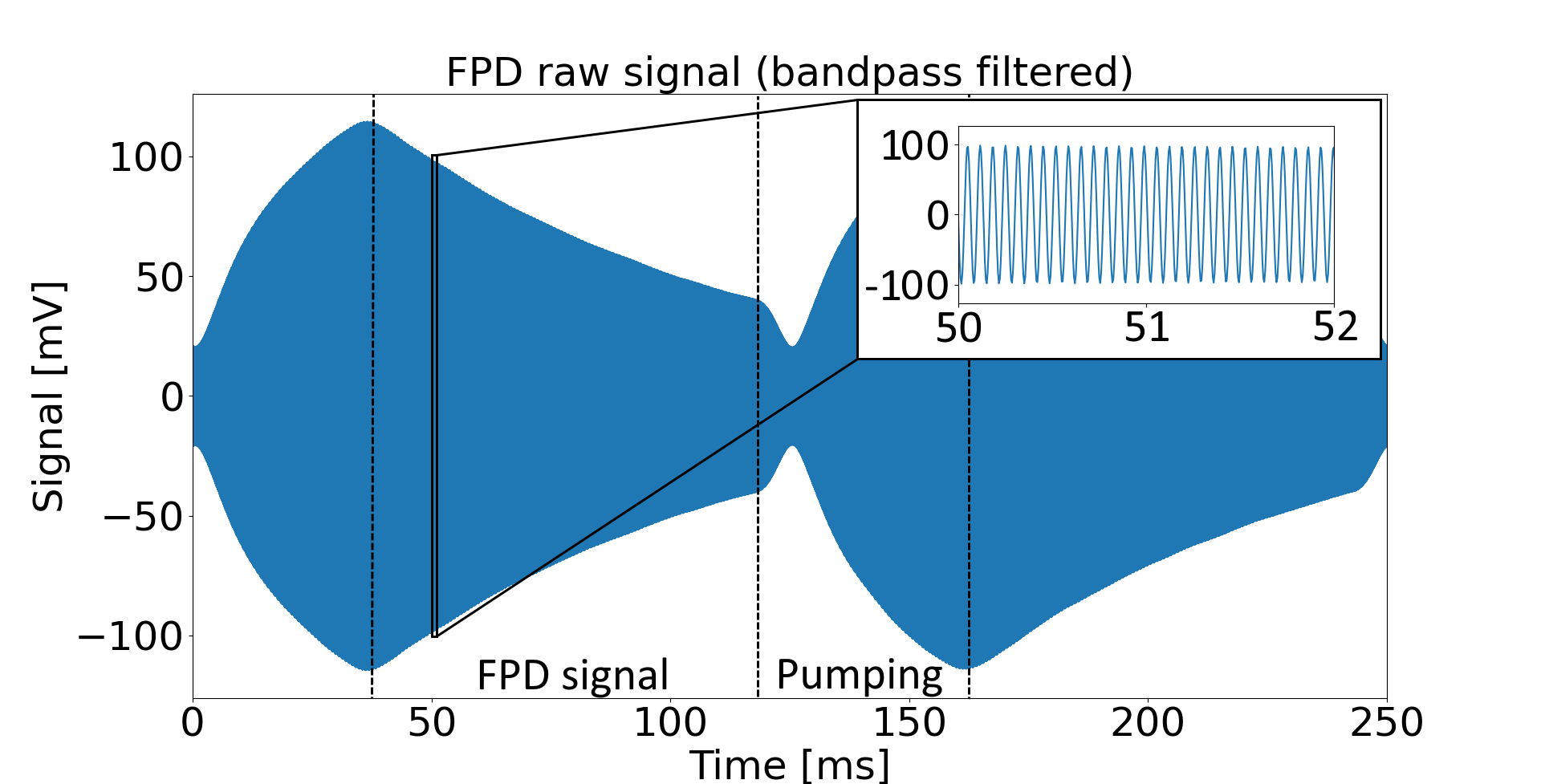}
	\caption{Example of a free precession decay dataset. The polarization of the Cs vapor builds up during an alignment pumping period of 40~ms. Afterwards spins precess freely while the polarization decays with a time constant of $T_2\approx 70\;\text{ms}$. Every 125~ms a magnetic field measurement is obtained from the decaying precession signal.}
	\label{fig:fpd}
\end{figure}
Two operation modes with different pumping schemes are used.

Forced oscillation scan (FO): In this mode the modulation frequency of the pump beam is swept over a range of $\omega_\text{L}\pm\SI{15}{\hertz}$ in 0.5~Hz increments to map the magnetic resonance response.
Simultaneously, the probe is analyzed using a lock-in amplifier referenced to the modulation frequency. 
The Larmor frequency $\omega_\text{L}$, together with amplitude and linewidth, is obtained from fitting the dataset to the resulting in-phase ("absorptive") and quadrature ("dispersive") components of a complex Lorentzian curve.\\
The bandwidth for determining magnetic fields from FO scans is limited to about 15~mHz due to the acquisition time of 60~s required to accumulate a full spectrum. 
Hence, FO mode is not suitable for monitoring fast magnetic field variations. Additionally, pump beam related systematic effects impair drift stability.
Free precession decay (FPD): After optical pumping at the resonance frequency $2\omega_\text{L}$, the probe beam's time-dependent polarization components are recorded for a duration of  $T_2 \approx \SI{70}{\milli\second}$, the time constant of free precession decay, allowing for higher bandwidth measurements.
A full measurement cycle, including pumping, takes about 125~ms, corresponding to a measurement bandwidth of 8~Hz (8 FPDs/s).\\
FPD mode is less sensitive to major systematic effects by separating pump and probe sequences. In particular, light-shift effects depending on pump light (frequency and power) and stray light reaching from the pump beam into the probe beam are suppressed.

Measurements were performed at PTB Berlin inside the magnetically shielded room BMSR-2\cite{bork_2000}, providing a low magnetic field environment with residual fields at the 0.5~nT level. A coil system mounted in the center of the shield (1.6~m diameter) provides a magnetic holding field of \SI{2.1}{\micro\tesla} with a measured gradient at the center of order \SI{0.5}{\nano\tesla\per\metre} over 12~cm. An active field stabilization mechanism using the signal from a LTc SQUID for feedback on correction coils is provided (PTB patent pending). The Cs sensors are aligned along the magnetic field and placed on a wooden table about 8~cm below the SQUID sensor.

A parameter scan was done for laser detuning, laser powers and duty cycle. The pump and probe beams are tuned to the $F=3\rightarrow F'$ and $F=4 \rightarrow F'$ transitions respectively. We use a probe power of typically $\SI{8}{\micro\watt}$ and a time-averaged power for the pump beam of $\SI{120}{\micro\watt}$. 
The duty cycle, i.e. pump pulse duration relative to the precession period, was adjusted to $\approx 20\%$.
In FPD mode we apply $\approx 600$ pump bursts corresponding to about 40~ms pumping duration (c.f. Fig. \ref{fig:fpd}).

For forced oscillation scans, the sensitivity is determined from the slope of the dispersive response in a resonance scan. 
Based on a dispersive slope of $\SI{35}{\milli\volt\per\hertz}$ and a noise floor of $\SI{9}{\micro\volt}$ (measured with the pump beam blocked) the zero-crossing uncertainty of the dispersive response is about \SI{37}{\femto\tesla} for a single FO scan of \SI{60}{\second}. 
Taking the lock-in amplifier ENBW\footnote{Equivalent noise bandwidth} of \SI{0.78}{\hertz} (integration time of \SI{100}{\milli\second} per data point at 24~dB$/$octave) into account, the magnetic field sensitivity is determined to be \SI{42}{\femto\tesla\per\sqrt{\hertz}}.
The order of frequencies during a FO scan is randomized to reduce the effect of systematic drifts, e.g. magnetic field variations or laser power drifts. 

\begin{figure}[t!]
	\centering
	\includegraphics[width=8.5cm]{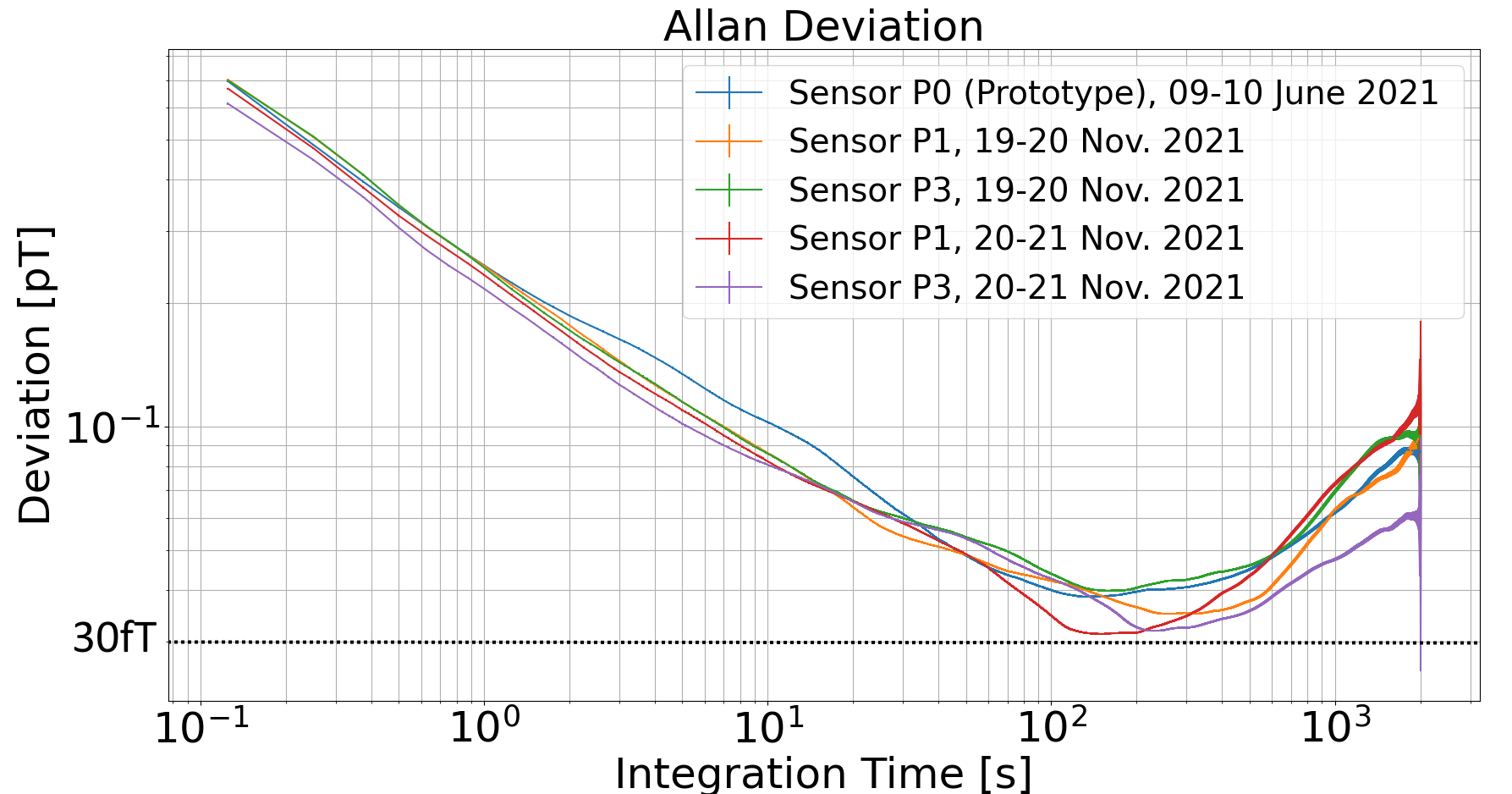}
	\caption{Allan standard deviation of measurements taken with different sensors on various days. The magnetic field was actively stabilized with feedback on the signal of a SQUID sensor. All measurements show a stability below 50~fT between 70 and 600~s.}
	\label{fig:fig2}
\end{figure}

In FPD mode, each dataset is sampled at 200~kHz, bandpass-filtered and fitted to an exponentially decaying sine wave to extract magnetic field information according to
\begin{equation}
	S= A\cdot \sin{(2\pi f_\text{R}t + \phi_0)}\cdot e^{-\lambda\cdot t} + c
\end{equation}
Here, $A$ denotes the signal amplitude, $f_\text{R}$ twice the Larmor frequency ($f_R = 2\gamma_\text{Cs}B$), $\phi_0$ accounts for an initial phase, $\lambda$ considers the decay constant of the FPD and $c$ an additional offset of the signal. 
Fig.~\ref{fig:fpd} shows an example dataset obtained in FPD mode, with build up of polarization during pumping and subsequent free precession in a 8 Hz periodic sequence.\\
An overlapping Allan deviation\cite{allan_1966} can be calculated to quantify the variation of the measured magnetic field on different time scales. 
The sensitivity for a single FPD is 600-700~fT, given as the average change between two consecutive FPDs.
From datasets of several thousand seconds recorded on different days using several sensors we demonstrate a reproducible, highly stable drift behavior of our sensor in FPD mode reaching a sensitivity of 35~fT within $200\;\text{s}$ and stability below $50\;\text{fT}$ between 70-600 s, shown in Fig.~\ref{fig:fig2}. \\
Limitations of the drift stability in our measurements beyond 300~s could originate from laser drifts or the active magnetic field stabilization based on the SQUID measurement. 

A distinction between magnetic field and sensor drift is difficult at the level of 35~fT. 
The laser power in the readout beam causes an observed light-shift in the magnetic field measurement of \SI{12}{\pico\tesla\per\micro\watt}. 
Based on measurements a conservative limit for the laser power stability is deduced resulting in an expected false effect of $<$10~fT for 500~s integration time.\\
The typically drift of the field stabilizing SQUID system of order pT/h indicates limitations from magnetic field drifts. 

By deploying several sensors of this type with comparable performance, we could demonstrate an upper limit for the magnetic field offset per sensor of less than 15~pT, as well as a linear response with a maximum deviation of linearity of $<50\;\text{fT}$ within a range of $100\;\text{pT}$\footnote{To be published.}. The magnetization of the sensor head was measured to be below 500~fT at 3~cm distance.

The theoretical sensitivity limit of the magnetic field sensor is based on the Cramer-Rao lower limit\cite{Gemmel2010} 
\begin{equation}
	\sigma_f \geq \frac{\sqrt{3}}{\pi}  \frac{\rho}{A}\cdot T^{-3/2} \sqrt{C(T,T_2^*)}
\end{equation}
With $C\approx 1.7$ for $T \approx T_2$, noise floor $\rho \approx 10\;\mathrm{\mu V/\sqrt{Hz}}$ and signal amplitude $A\approx 100\;\text{mV}$
this yields $\sigma_f \approx 550\;\text{fT}$ in good agreement with the measured single FPD variation of 600-700~fT.
In summary, we have demonstrated a highly stable NMOR-based atomic magnetometer with its sensitivity optimized for long integration times and low systematic drifts.
The modular, screw-mounted design allows easy maintenance and is remotely operated via 8~m long fibers. The sensor head is fully optical, non-magnetic and non-metallic, while being robust in performance and scalable to an array of magnetometers.\\
In FPD mode the sensitivity has been demonstrated to be as low as 35~fT for 200~s integration time with stability below 50~fT for long time scales of up to 600 s. Our measurements are systematically very clean due to separation of pump and probe sequences and using linear polarized light.
Our results indicate that the performance of the device at $>$100~s integration time is limited by drifts of the actively stabilized magnetic field. \\
The sensor fulfills the performance requirements as a reference magnetometer system to monitor the stability of the magnetic field distribution in new neutron EDM searches, as well as other high precision experiments relying on highly stable magnetic field environments.
\bigskip

This research was supported by the Excellence Cluster ORIGINS which is funded by the Deutsche Forschungsgemeinschaft (DFG, German Research Foundation) under Germany's Excellence Strategy – EXC-2094 and the DFG projects FI-1663-5 and FI-1663-12. Support was also received in part by National Science Foundation Grant PHY-1812377 and the National Natural Science Foundation of China (grant number 51861135308). This research was further supported by DFG projects 410292433 and 237136261. \\ We acknowledge the support of the Core Facility “Metrology of ultra-low magnetic fields” at Physikalisch-Technische Bundesanstalt which receives funding from the Deutsche Forschungsgemeinschaft – DFG (funding code DFG KO 5321/3-1 and TR 408/11-1). We thank the PTB team at BMSR-2, especially Allard Schnabel, for providing and optimizing the magnetic field stabilization for our task.\\ DB thanks Caltech, under the Moore Scholars program, where some of this work was done.
\bigskip

The data that support the findings of this study are available from the corresponding author upon reasonable request.
\bigskip

\bibliography{2021_CesiumPaper}
\end{document}